\documentclass[12pt]{article}
\usepackage{Input/spie,graphicx,xspace,amsmath,amssymb}
\graphicspath{{figures_changepoint/}} 
\input{Input/alphabet}   
\input{Input/abrege}     
\def\pth#1{\left(#1\right)}                
\def\acc#1{\left\{#1\right\}}              
\def\cro#1{\left[#1\right]}

\def\th{{\mathrm{th}}}                

\newsavebox{\fminibox}
\newlength{\fminilength}
\newenvironment{fminipage}[1][\linewidth]
  {\setlength{\fminilength}{#1}
   \begin{lrbox}{\fminibox}\begin{minipage}{\fminilength}}
  {\end{minipage}\end{lrbox}\noindent\fbox{\usebox{\fminibox}}}

  \def\+{^\dagger}

\def\nequiv{\not\kern-.05em\equiv}
\def\egal{\kern-.5em=\kern-.5em}        
\def\propt{\kern-.2em\propto\kern-.2em} 

\def\argmax{\mathop{\mathrm{arg\,max}}} 
\def\argmin{\mathop{\mathrm{arg\,min}}} 
\def\intdouble{\int\kern-0.3em\int}
\def\inttriple{\int\kern-0.3em\int\kern-0.3em\int}

\def\rond#1{\overset{\kern-0.33em~_\circ}{#1}}
\def\rondit[#1]#2{\overset{\kern#1~_\circ}{#2}}

\def\babs{\begin{abstract}}             \def\eabs{\end{abstract}}
\def\barr{\begin{array}}                \def\earr{\end{array}}
\def\bcc{\begin{center}}                \def\ecc{\end{center}}

\def\bdes{\begin{description}}          \def\edes{\end{description}}
\def\bdoc{
\begin{document}}             \def\edoc{\end{document}}
\def\ben{\begin{enumerate}}             \def\een{\end{enumerate}}
\def\beqn{\begin{eqnarray}}             \def\eeqn{\end{eqnarray}}
\def\beqnl#1{\beqn\label{#1}}           \def\eeqnl#1{\label{#1}\eeqn}
\def\beqnx{\begin{eqnarray*}}           \def\eeqnx{\end{eqnarray*}}
\def\bseqn{\begin{subeqnarray}}         \def\eseqn{\end{subeqnarray}}
\def\beq#1\eeq{\begin{equation}#1\end{equation}}
\def\bal#1\eal{\begin{align}#1\end{align}}
\def\balx#1\ealx{\begin{align*}#1\end{align*}}
\def\beqx{$$}                           \def\eeqx{$$}
\def\bfig{\protect\begin{figure}}       \def\efig{\protect\end{figure}}
\def\bfigx{\protect\begin{figure*}}     \def\efigx{\protect\end{figure*}}
\def\bfigt{\protect\begin{figurette}}   \def\efigt{\protect\end{figurette}}
\def\bfl{\begin{flushleft}}             \def\efl{\end{flushleft}}
\def\bfr{\begin{flushright}}            \def\efr{\end{flushright}}
\def\bit{\begin{itemize}}               \def\eit{\end{itemize}}
\def\bmi{\begin{minipage}}              \def\emi{\end{minipage}}
\def\bfmi{\begin{fminipage}}            \def\efmi{\end{fminipage}}
\def\bpic{\begin{picture}}              \def\epic{\end{picture}}
\def\bqu{\begin{quote}}                 \def\equ{\end{quote}}
\def\bqun{\begin{quotation}}            \def\equn{\end{quotation}}
\def\bsl{\begin{slide}}                 \def\esl{\end{slide}}
\def\btabb{\begin{tabbing}}             \def\etabb{\end{tabbing}}
\def\btabl{\begin{table}}               \def\etabl{\end{table}}
\def\btablx{\begin{table*}}             \def\etablx{\end{table*}}
\def\btab{\begin{tabular}} 
\def\btabu{\begin{tabular}}             \def\etabu{\end{tabular}}
\def\btabx{\begin{tabular*}}            \def\etabx{\end{tabular*}}
\def\bbib{\begin{thebibliography}{999}} \def\ebib{\end{thebibliography}}
\def\bver{\begin{verbatim}}             \def\ever{\end{verbatim}}
\def\bca{\begin{cases}}                  \def\eca{\end{cases}}
   
\def\bm#1{\mbox{\boldmath #1}}
\def\d#1{\,\hbox{d}#1}
\def\bs{\bar{s}}
\def\us{\underline{s}}
\def\fmin{f_{\mbox{min}}}
\def\fmax{f_{\mbox{max}}}
\def\expf#1{\mbox{exp}\left\{#1\right\}}
\def\argmin#1#2{\mbox{arg}\min_{#1}\left\{#2\right\}}
\def\argmax#1#2{\mbox{arg}\max_{#1}\left\{#2\right\}}

\def\lra{\longrightarrow}
\def\fh{\widehat{f}}
\def\fbh{\widehat{\fb}}
\def\Rbh{\widehat{\Rb}}
\def\Sigmabh{\widehat{\Sigmab}}

\def\rbh{\widehat{\rb}}
\def\qbh{\widehat{\qb}}
\def\xbh{\widehat{\xb}}
\def\tbh{\widehat{\tb}}
\def\thetabh{\widehat{\thetab}}

\def\disp{\displaystyle}
\def\vsm{\vspace*{-12pt}}
\def\hsm{\hspace*{-3em}}

\def\pmata#1#2{\left(\barr{c} #1 \\ #2 \earr\right)}
\def\pmatb#1#2#3#4{\left(\barr{cc} #1 & #2 \\ #3 & #4 \earr\right)}

\def\th{\widehat{t}}
\def\xh{\widehat{x}}
\def\lambdah{\widehat{\lambda}}
\def\muh{\widehat{\mu}}
\def\sigmah{\widehat{\sigma}}
\def\rhoh{\widehat{\rho}}
\def\betah{\widehat{\beta}}
\def\alphah{\widehat{\alpha}}
\def\tbh{\widehat{\tb}}
\def\mubh{\widehat{\mub}}
\def\sigmabh{\widehat{\sigmab}}
\def\rhobh{\widehat{\rhob}}
\def\oneb{\mbox{\bf 1}}

\title{A Bayesian approach to change point analysis of discrete time series}

\author{Ali Mohammad-Djafari and Olivier F\'eron\\[.4cm]
  Laboratoire des Signaux et Syst\`emes,\\  
  Unit\'e mixte de recherche 8506 (CNRS-Sup\'elec-UPS) \\  
  Sup\'elec, Plateau de Moulon, 91192 Gif-sur-Yvette, France\\ 
  emails = {djafari,feron@lss.supelec.fr} 
}
\authorinfo{\noindent \hspace*{-0.5cm}${}^\ast$Correspondence:~E-mail:
djafari@lss.supelec.fr}
\date{}

\bdoc
\maketitle

\begin{abstract}
In this work we consider time series with a finite number of discrete point changes. We assume that the data in each segment follows a different probability density functions (pdf). 
We focus on the case where the data in all segments are modeled by Gaussian probability density functions with different means, variances and correlation lengths. 
We put a prior law  on the change point instances (Poisson process) as well as on these different parameters(conjugate priors) and give the expression of the posterior probality distributions of these change points. The computations are done by using an appropriate Markov Chain Monte Carlo (MCMC) technique. 

The problem as we stated can also be considered as an unsupervised classification and/or segmentation of the time serie. 
This analogy gives us the possibility to propose alternative modeling and computation of change points, which are more appropriate for multivariate signals, for example in image processing.
\\ ~\\ 
{\bf key words:}~ Bayesian change-points estimation, classification and segmentation. 
\end{abstract}

\section{Introduction}

Figure 1 shows typical change point problems we consider in this work. 
Note that, very often people consider problems in which there is only one change point \cite{Basseville88}. Here we propose to consider more general problems with any number of change points. However, very often the change point analysis problems need online or real time detection algorithms \cite{Wax91,Kormylo82,Chi85,Goutsias88}, while here, we focus only on off line methods where we assume that we have gathered all the data and we want to analyse it to detect change points who have been occured during the observation time. 
Also, even if we consider here change point estimation of 1-D time series, we can extend the proposed method to multivariate data, for example the images where the change point problems become equivalent to segmentation. 
One more point to position this work is that, very often the models used in change point problems assume to know perfectly the model of the signal in each segment, \ie a linear or nonlinear regression model \cite{Goutsias88,Oliver96,Hughes99,Fitzgibbon00,Fitzgibbon02}, while here, we use a probabilistic model for the signals in each segment which gives probably more generality and applicability when we do not know perfectly those models.

\bfig[hbt]
\bcc
\includegraphics[width=150mm,height=75mm]{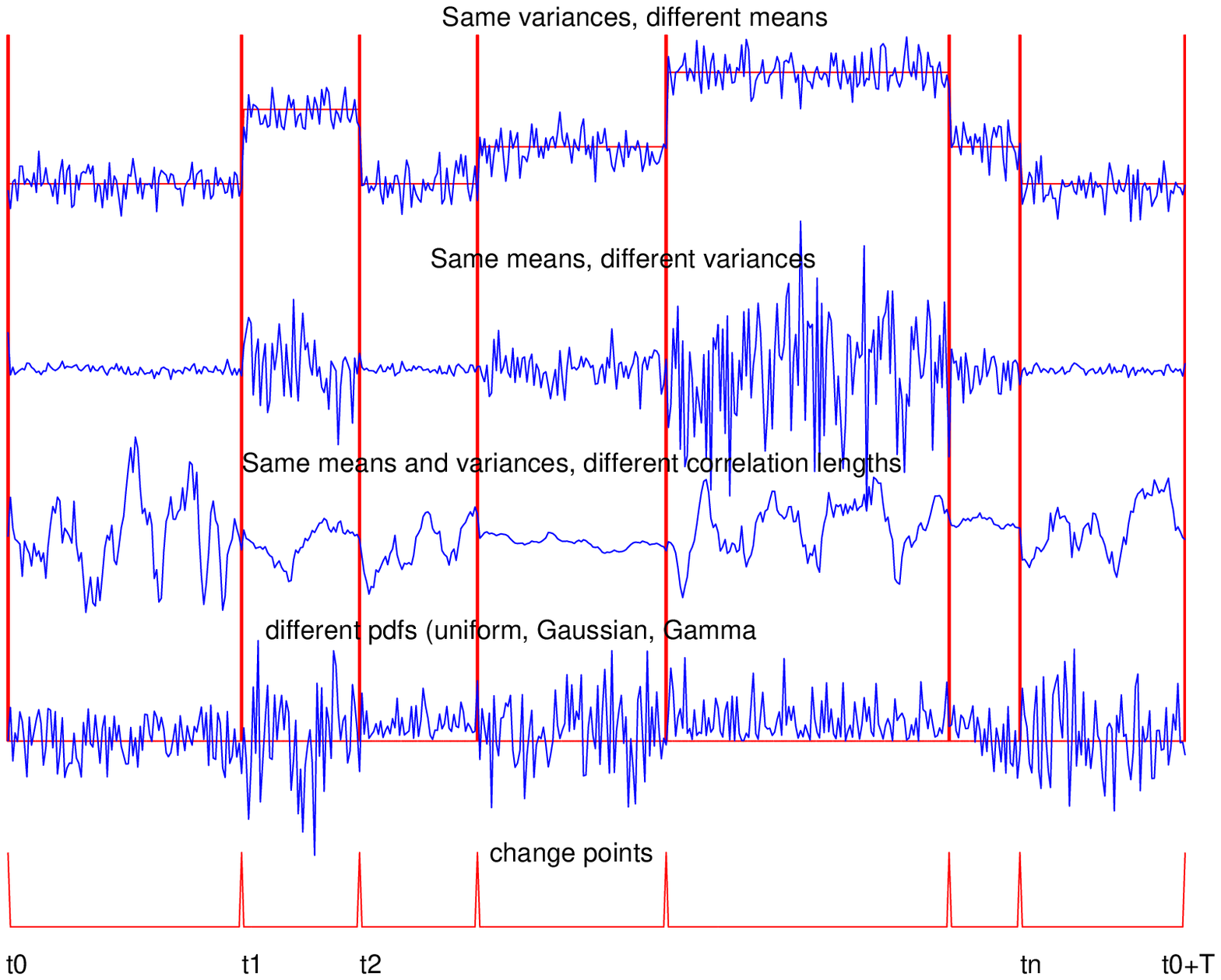} 
\ecc
\caption[Change point problems description.]{Change point problems description: 
In the first row, only mean values of the different segments are different. In the second row, only variances are changed. In the third row only the correlation strengths are changed. In the fifth row, the whole nature shape of their probability distribution have been changed. The last row show the change points $t_n$.}
\label{fig1}
\efig

More specifically, we model the time series by a hierarchical Gauss-Markov modeling with hidden varaibles which are themselves modeled by a Markov model. Though, in each segment which corresponds to a particular value of the hidden variable, the time series is assumed to be modeled by a stationnary Gauss-Markov model. However, we choosed a simple parametric model defined only with three parameters of mean $\mu$, variance $\sigma^2=1/\tau$ and a parameter $\rho$ measuring the local 
correlation strength of the neighboring samples. 

The choice of the hidden variable is also important. We have studied three different modeling: i) change point time instants $t_n$, ii) classification labels $z_n$ 
or iii) a Bernouilli variable $q_n$ which is always equal to zero except when a change point occurs. 

The rest of the paper is organized as follows: In the next section we introduce the notations and fixe the objectives of the paper. In section 3 we consider the model with explicite change point times as the hidden variables and propose particular modeling for them and an MCMC algorithm to compute their \apost probabilities. In sections 4 and 5 we consider the two other aformentionned models. Finally, we show some simulation results and present our conclusions and perspectives.


\newpage
\section{Notations and modeling}
We note by $\xb=[x(t_0), \cdots, x(t_0+T)]'$ the vector containing the data observed from time $t_0$ to $t_0+T$. We note by 
$\tb=[t_1,\cdots,t_N]'$ the unknown change points and note  
$\xb=[\xb_0, \xb_1, \cdots , \xb_N]'$ 
where $\xb_n=[x(t_n), x(t_n+1), \cdots, x(t_{n+1})]', \quad n=0,\cdots,N$ 
represent the data samples in each segment. In the following we will have $t_{N+1}=T$.

We model the data 
$\xb_n=[x(t_n), x(t_n+1), \cdots, x(t_{n+1})]', \quad n=0,\cdots,N$ 
in each segment by a Gauss-Markov chain:

\beqn
p(x(t_n))&=&\Nc(\mu_n, \sigma_n^2) \nonumber \\ 
p(x(t_n+l)|x(t_n+l-1))&=&\Nc(\rho_n \, x(t_n+l-1)+(1-\rho_n)\mu_n, \sigma_n^2(1-\rho_n^2)), 
\quad l=1,\cdots,l_n -1 \nonumber \\ 
\mbox{with~~~}&& 
l_n=t_{n+1}-t_n+1=\dim{[\xb_n]}  
\label{eq1}
\eeqn
Then we have 
\beqn
p(\xb_n) &=&p(x(t_n)) \prod_{l=1}^{l_n} p(x(t_n+l)|x(t_n+l-1)) \nonumber \\ 
p(\xb_n) &\propto& \expf{-\frac{1}{2\sigma_n^2} (x(t_n)-\mu_n)^2} \nonumber\\ 
         &&\expf{-\frac{1}{2(\sigma_n^2(1-\rho_n^2))} \sum_{l=1}^{l_n} 
[x(t_n+l)- \rho_n x(t_n+l-1)-(1-\rho_n)\mu_n]^2} \nonumber\\ 
p(\xb_n)&=&\Nc(\mu_n\oneb, \Sigma_n) 
\mbox{~~with~~} \Sigmab_n=\sigma_n^2 \, 
\mbox{Toeplitz}([1, \rho_n, \rho_n^2,\cdots, \rho_n^{l_n}])
\label{eq2}
\eeqn

Noting by $\tb=[t_1,\cdots,t_N]$ the vector of the change points and 
assuming that the samples from any two segments are independent, we can write:
\beq \label{eq3}
p(\xb|\tb,\thetab, N)=\prod_{n=0}^N \Nc(\mu_n\oneb, \Sigmab_n)
=\pth{\prod_{n=0}^N \frac{|\Sigmab_n|^{-1/2}}{(2\pi)^{(l_n/2)}}}
\expf{-\frac{1}{2} \sum_{n=0}^N  (\xb_n-\mu_n\oneb)'\Sigmab_n^{-1}(\xb_n-\mu_n\oneb)}
\eeq
where we noted   
$\thetab=\acc{\mu_n,\sigma_n,\rho_n,\; n=0,\cdots,N}$. 

Note that 
\beq \label{eq4}
-\ln p(\xb|\tb,\thetab,N)
=\sum_{n=0}^N (l_n/2) \ln (2\pi)
+\frac{1}{2} \sum_{n=0}^N \ln {|\Sigmab_n|}
-\frac{1}{2} \sum_{n=0}^N  (\xb_n-\mu_n\oneb)'\Sigmab_n^{-1}(\xb_n-\mu_n\oneb)
\eeq
and when the data are \iid, ($\Sigmab_n=\sigma_n\Ib$) this becomes 
\beq \label{eq5}
-\ln p(\xb|\tb,\thetab,N)
=(T/2)\ln (2\pi)
+\sum_{n=0}^N (l_n/2) \ln {\sigma_n^2}
- \sum_{n=0}^N  \frac{\|(\xb_n-\mu_n\oneb)\|^2}{2\sigma_n^2} 
\eeq

Then, the inference problems we will be faced are the following:
\ben
\item Infer on $\thetab$ given $\xb$ and $\tb$;
\item Infer on $\tb$ given  $\xb$ and $\thetab$;
\item Infer on $\tb$ and $\thetab$ given $\xb$;
\item Infer on $\thetab$ given $\xb$.  
\item Infer on $\tb$ given  $\xb$; 
\een
It is clear that the first problem is the easiest. 

The classical maximum likelihood estimation (MLE) approach can handle only the 
first three problems by maximizing $p(\xb|\tb,\thetab)$, respectively, with respect to $\thetab$, to $\tb$ and jointly $(\tbh,\thetabh)$: 

\bit
\item Estimating $\thetab$ given $\xb$ and $\tb$: \quad 
\(
\thetabh=\argmax{\thetab}{p(\xb|\tb,\thetab)}
\)

\item Estimating $\tb$ given  $\xb$ and $\thetab$: \quad 
\(
\tbh=\argmax{\tb}{p(\xb|\tb,\thetab)}
\)

\item Estimating $\tb$ and $\thetab$ given $\xb$: \quad 
\(
(\tbh,\thetabh)=\argmax{(\tb,\thetab)}{p(\xb|\tb,\thetab)}
\)
\eit
However, we must be careful to check the boundedness of the likelihood function before using any optimization algorithm. 
The optimization with respect to $\thetab$ when $\tb$ is known can be done easily, but the optimization with respect to $\tb$ is very hard and computationally costly. 

The two last problems cannot be handled easily because they need to define the likelihood fuctions $p(\xb|\thetab)$ and $p(\xb|\tb)$ which need integrations with respect to $\tb$ or $\thetab$ of $p(\xb|\tb,\thetab)$. There may not be possible to find analytical expressions for these integrals which may even not exist.


\section{Bayesian estimation of the change point time instants}

In Bayesian approach, one assigns prior probability laws on both $\tb$ and $\thetab$ and use the posterior probability law $p(\tb,\thetab|\xb)$ as a tool for doing any inference. Choosing a prior pdf for $\tb$ is also usual in classical approach. A simple model is the following:
\beq \label{eq6}
t_n=t_{n-1}+\epsilon_n \quad \mbox{with}\quad \epsilon_n\sim\Pc(\lambda),
\eeq

\noindent
where $\varepsilon_n$ are assumed iid end $\lambda$ is the \aprio mean value of time intervals $(t_n-t_{n-1})$. if $N$ is the number of changepoint we can take $\lambda=\frac{T}{N+1}$. With this modeling we have :
\rem{
can either be a constant or equal to the mean value of the past intervals:  
$\lambda_n=\frac{1}{n-1}\sum_{k=1}^n (t_n-t_{n-1}), \quad n>2$
}
\beq \label{eq7}
\barr{l}
p(\tb|\lambda)=\prod_{n=1}^{N+1} \Pc(t_n-t_{n-1}|\lambda)=
\prod_{n=1}^{N+1} e^{-\lambda} \frac{\lambda^{(t_n-t_{n-1})}}{(t_n-t_{n-1})!}
\\
\ln p(\tb|\lambda)=
 -(N+1)\lambda + \ln(\lambda)\sum_{n=1}^{N+1} (t_n-t_{n-1})-\sum_{n=1}^{N+1} \ln((t_n-t_{n-1})!)%
\earr
\eeq


With this prior selection, we have
\beq
p(\xb,\tb|\thetab,N)=p(\xb|\tb,\thetab,N) \, p(\tb|\lambda,N)
\eeq
and
\beq \label{eq9}
p(\tb|\xb,\thetab,N)\propto p(\xb|\tb,\thetab,N) \, p(\tb|\lambda,N)
\eeq

In Bayesian approach, one goes one step further with assigning prior probability laws to the hyperparameters $\thetab$, 
\ie $p(\thetab)$ 
and then one writes the joint \apost:

\beq
p(\tb,\thetab|\xb,\lambda,N) \propto p(\xb|\tb,\thetab,N)\, p(\tb|\lambda,N) \, p(\thetab|N)   
\eeq
where here we noted   
$\thetab=\acc{\mu_n,\sigma^2_n,\rho_n,\; n=1,\cdots,N}$. 

To go further in details, we need to assign $p(\thetab)$.
 The following is our selection:
\beqnx
p(\mu_n) &=& \Nc(\mu_0,\sigma_0^2)\\ 
p(\sigma_n^2) &=& {\cal IG}(\alpha_0,\beta_0) \\ 
p(\rho_n) &=& \Uc([0,1]) 
\eeqnx 
which correspond mainely to the conjugate or reference priors. 

Given all these, we propose the following Gibbs MCMC algorithm:
\[
\barr{lllll}
\mbox{Iterate until convergency} \\ 
\mbox{.~~sample~~} \tb &\mbox{using}& p(\tb|\xb,\thetab,N)  
\\ 
\mbox{.~~sample~~} \theta_n : \\
\qquad \mu_n &\mbox{using}& p(\mu_n|\xb,\tb,N) \\ 
\qquad \sigma_n^2 &\mbox{using}& p(\sigma_n^2|\xb,\tb,N) \\ 
\qquad \rho_n &\mbox{using}& p(\rho_n|\xb,\tb,N) \\ 
\earr
\]

\subsection{Sampling $\tb$ using $p(\tb|\xb,\thetab,N)$}
P. Fearnhead showed \cite{Fearnhead} that it is possible to perform perfect simulation of $p(\tb|\xb,\thetab,N)$ when we have assumed that segments of data separated by a changepoint $t_n$ are independant. This simulation can be obtained by a method based on recursion on the changepoints. An approximation of this method is possible to obtain an algorithm whose computational cost is linear in the number of observations. The main principle of this algorithm is to compute the following probabilities :\\
\noindent
Let note $\xb_{t:s}=[x(t),x(t+1),\dots,x(s)]$, and
\begin{eqnarray*}
R(t,s|\lambda) & = & p(\xb_{t:s}|t,s \mbox{ in the same segment},\lambda) \\
Q(t|\lambda) & = & p(\xb_{t:s}|\mbox{ changepoint at } t-1,\lambda), \quad Q(1)=p(\xb|\lambda) 
\end{eqnarray*} 
\noindent
Let also note $F(t|\lambda)$ the associated cumulative distribution function of the prior density $\mathcal P(t_n-t_{n-1}|\lambda)$ which is defined by (7). \\
\noindent
We compute $R(t,s|\lambda)$ with the following relation :
\beq
R(t,s)|\lambda)=\int p(\xb_{t:s}|\thetab,\lambda)p(\thetab) d\thetab
\nonumber
\eeq 
The computation of $Q(t|\lambda)$ can be done recursively by the following result : for $t=1,\dots,T$,
\beq
Q(t|\lambda)=\sum^{T-1}_{s=t}R(t,s|\lambda)Q(s+1|\lambda)\mathcal P(s+1-t|\lambda)+R(t,T|\lambda)(1-F(T-t|\lambda)),
\nonumber
\eeq
\noindent
This result is shown by P. Fearnhead \cite{Fearnhead} . And he also demonstrates that the posterior distribution of $t_n$ given $t_{n-1}$ is
\begin{eqnarray*}
p(t_n|t_{n-1},\xb,\lambda)=\frac{R(t_{n-1},t_n|\lambda)Q(t_n+1|\lambda) \mathcal P(t_n-t_{n-1}|\lambda)}{Q(t_{n-1}|\lambda)}
\end{eqnarray*}
\noindent
and the posterior distribution of no further changepoint is given by
\beq
p(t_n=T|t_{n-1},\xb,\lambda)=\frac{R(t_{n-1},T|\lambda)(1-F(T-t_{n-1}-1|\lambda))}{Q(t_{n-1}|\lambda)}
\nonumber
\eeq

\subsection{Sampling $\theta_n$ using $p(\theta_n|\xb,\tb,N)$}
We may note that, thanks to the conjugacy, we have: 
\beqnx
p(\mu_n|\xb,\tb)&=&\Nc(\muh_n,\sigmah_n^2) \mbox{~~with~~} 
\left\{\barr{l}
\muh_n= \sigmah_n^2 \left[ \frac{\mu_0}{\sigma_0^2}+\oneb'\Sigmab_n^{-1} \xb_n \right]\\ 
\sigmah_n^2= \left( \oneb'\Sigmab_n^{-1}\oneb + \frac{1}{\sigma_0^2} \right)^{-1}
\earr\right. 
\\ 
p(\sigma_n^2|\xb,\tb)&=&{\cal IG}(\alphah_n,\betah_n)  \mbox{~~with~~}  
\left\{\barr{l}
\alphah_n= \alpha_0 + \frac{l_n}{2} \\ 
\betah_n= \beta_0 + \frac{1}{2}(\xb_n-\mu_n\oneb)'\Rb_n^{-1}(\xb_n-\mu_n\oneb),
\earr\right. 
\eeqnx
\noindent
where $\Rb_n = \mbox{Toeplitz}([1, \rho_n, \rho_n^2,\cdots, \rho_n^{l_n}])$. Then the simulation of these densities is quite simple.\\  \\ \\
\noindent
$p(\rho_n|\xb,\tb)$ is not a classical law. Its expression is given by : 
\begin{eqnarray*}
p(\rho_n|\xb,\tb,N) & = & \prod_{n=0}^N p(\rho_n|\xb_n,\tb,N) \\
& \propto & \left( \frac{1}{\sigma_n^2(1-\rho_n^2)}\right)^{\frac{ln}{2}} \exp \left\{- \frac{1}{2\sigma_n^2(1-\rho_n^2)} (\xb_n-\mu_n\oneb)'\Rb_n^{-1}(\xb_n-\mu_n\oneb) \right\} \\
& \propto & \left( \frac{1}{\sigma_n^2(1-\rho_n^2)}\right)^{\frac{ln}{2}} \exp \left\{- \frac{1}{2\sigma_n^2(1-\rho_n^2)} \sum_{l=1}^{ln} (x(t_n+l)-\rho_n x(t_n+l-1)-(1-\rho_n)\mu_n)^2 \right\}
\end{eqnarray*}
\noindent
Then we can not sample easily this density. \\
\noindent
The solution we propose is to use, in this step, a Hastings-Metropolis algorithm for sampling this density. As an instrumental density we propose to use a Gaussian approximation of the posterior density, \ie we estimate the mean $m_{\rho_n}$ and the variance $\sigma^2_{\rho_n}$ of $p(\rho_n|\xb,\tb,N)$ and we use a Gaussian law $\mathcal N(m_{\rho_n},\sigma^2_{\rho_n})$ to obtain a sample. This sample is accepted or rejected following $p(\rho_n|\xb,\tb,N)$. In practice we compute $m_{\rho_n}$ and $\sigma^2_{\rho_n}$ calculating by approximation of their definition :
\begin{eqnarray*}
m_{\rho_n} & \lra & \int_0^1 \rho_n \quad p(\rho_n|\xb,\tb,N) \\
\sigma^2_{\rho_n} & \lra & \int_0^1 \rho_n^2 \quad p(\rho_n|\xb,\tb,N) - m_{\rho_n}^2
\end{eqnarray*} 
\noindent


\newpage
\section{Other formulations}
Other formulation can also exist. 
We introduce two sets of hidden variables 
 
\centerline{
$\zb=[z(t_0), \cdots, z(t_0+T)]'$ and  
$\qb=[q(t_0), \cdots, q(t_0+T)]'$ 
}
where 
\beq
\barr{l}
q(t)=\left\{\barr{ll} 
1 & \mbox{if~}  z(t)\not= z(t-1) \\ 
0 & \mbox{elsewhere} 
\earr\right.
=\left\{\barr{ll} 
1 & \mbox{if~}  t=t_n, n=0,\cdots,N \\ 
0 & \mbox{elsewhere} 
\earr\right.
\earr.
\eeq
and where $z(t)$ takes an integer value $k$ in each segment : $k=1,\dots,N+1$. With these two related hidden variables, we can propose two other modeling to be used in change point analysis. For example, $\qb$ can be modeled by a Bernouilli process
\[
P(\Qb=\qb)=\lambda^{\sum_j q_j} (1-\lambda)^{\sum_j (1-q_j)}
=\lambda^{\sum_j q_j} (1-\lambda)^{N -\sum_j q_j}
\]
and $\zb$ can be modeled by a Mrkov chain, \ie 
$\{z(t), t=1,\cdots,T\}$ forms a Markov chain:
\beq
\barr{l}
P(z(t)=k)=p_k, \quad k=1,\cdots,K,\\ 
P(z(t)=k|z(t-1)=l)=p_{kl}, \quad\mbox{with~~} \sum_k p_{kl}=1. 
\earr
\eeq
These two models are related. In the first one, $\lambda$ plays the role of the mean value of the segment lengths and in the second $p_k$ and $p_{kl}$ give more precise control of the segment lengths.
In the multivariate case, or more precisely in  bivariate case (image processing), $\qb$ may represent the contours and $\zb$ the labels for the regions in the image. 
Then, we may also give a Markov model for them. For example, if we note by 
$r\in \Sc$ the position of a pixel, $\Sc$ the set of pixels positions and 
by $\Vc(r)$ the set of pixels in the neighorhood of the pixel position $r$, 
we may use an Ising model for $\qb$
\beq
P(\Qb=\qb)\propto \expf{-\rho \sum_{r\in\Sc} \sum_{s\in\Vc(r)} \delta(z(r)-z(s))}
\eeq
or a Potts model for $\zb$: 
\beq
P(\zb)\propto \expf{-\rho \sum_{r\in\Sc} \sum_{s\in\Vc(r)} 
\delta(z(r)-z(s))}. 
\eeq
where $rho$ in the first controls the mean lengths of the contours in the image and in the second the mean surface of the regions in the image.
Other more complexe modelings are also possible. 

With these auxiliary variables, we can write
\beq
p(\xb|\zb,\thetab)=\sum_{n=1}^N P(z_j=n) \Nc(\mu_n\oneb, \Sigmab_n)
=\sum_{n=1}^N p_k \Nc(\mu_n\oneb, \Sigmab_n)
\eeq
if we choose $K=N$. Here, 
$\thetab=\acc{N,\acc{\mu_n,\sigma_n, p_n,\; n=1,\cdots,N}, \pth{p_{kl}, \; k,l=1,\cdots,N}}$ and the model is a mixture of Gaussians. 

We can again assign appropriate prior law on $\thetab$ and give the expression of $p(\zb,\thetab|\xb)$ and do any inference on $\zb$, $\thetab$. 

Finally, we can also use $\qb$ as the auxiliary variable and write
\beqn
p(\xb|\qb,\thetab)&=&
(2\pi)^{-N/2} 
\pth{\prod_{n=1}^N 1/\sigma_n} 
\expf{-\frac{1}{2\sigma_n^2} \sum_{n=1}^N \pth{x(t_n)-\mu_n}^2} \nonumber \\ 
&+&
(2\pi)^{-(T-N)/2} 
\pth{\prod_{n=1}^N 1/\sigma_n^{(l_n-1)}}
\expf{-\frac{1}{2\sigma_n^2} \sum_{j=1}^T 
(1-q_j) \pth{x_{j}-x_{j-1}}^2} \nonumber \\ 
&=&
(2\pi)^{-T/2} 
\pth{\prod_{n=1}^N 1/\sigma_n^{(l_n)}}
\expf{-\frac{1}{2\sigma_n^2} \sum_{j=1}^T 
\cro{(1-q_j) \pth{x_{j}-x_{j-1}}^2
+q_j\pth{x_{j}-\mu_n}}} \nonumber \\ 
\eeqn
and again assign appropriate prior law on $\thetab$ and give the expression of $p(\qb,\thetab|\xb)$ and do any inference on $\qb$, $\thetab$. We are still working on using these auxiliary hidden variables particularly for applications in data fusion in image processing and we will report on these works very soon.


\newpage
\section{Simulation results}
To test the feasability and to mesaure the performances of the proposed algorithms, we generated a few simple cases corresponding to only changes of one of the three parameters $\mu_n$, $\sigma^2_n$ and $\rho_n$. \\
\noindent
In each case we present the data, the histogram of the \apost samples of $\tb$ during the first and the last iterations of the MCMC algorithm. For each case we also give the value of the parameters used to simulate the data, the estimated values when the changepoints are known and the estimated values by the proposed method.

\newpage
\subsection{Change of the means}

We can see in figure \ref{fig_mean} that we obtain precise results on the position of the changepoints. In the case of change of means, the algorithm is very fast to converge to the good solution. In fact it needs only few iterations (about 5). The main cause of this results is the importance of the means in the likelihood $p(\xb|\tb,\thetab,N)$. \\
We can also see in table 1 that the estimations of the means are very precise, particularly when the size of the segment is long.
\bfig[hbt]
\bcc
\includegraphics[width=150mm,height=75mm]{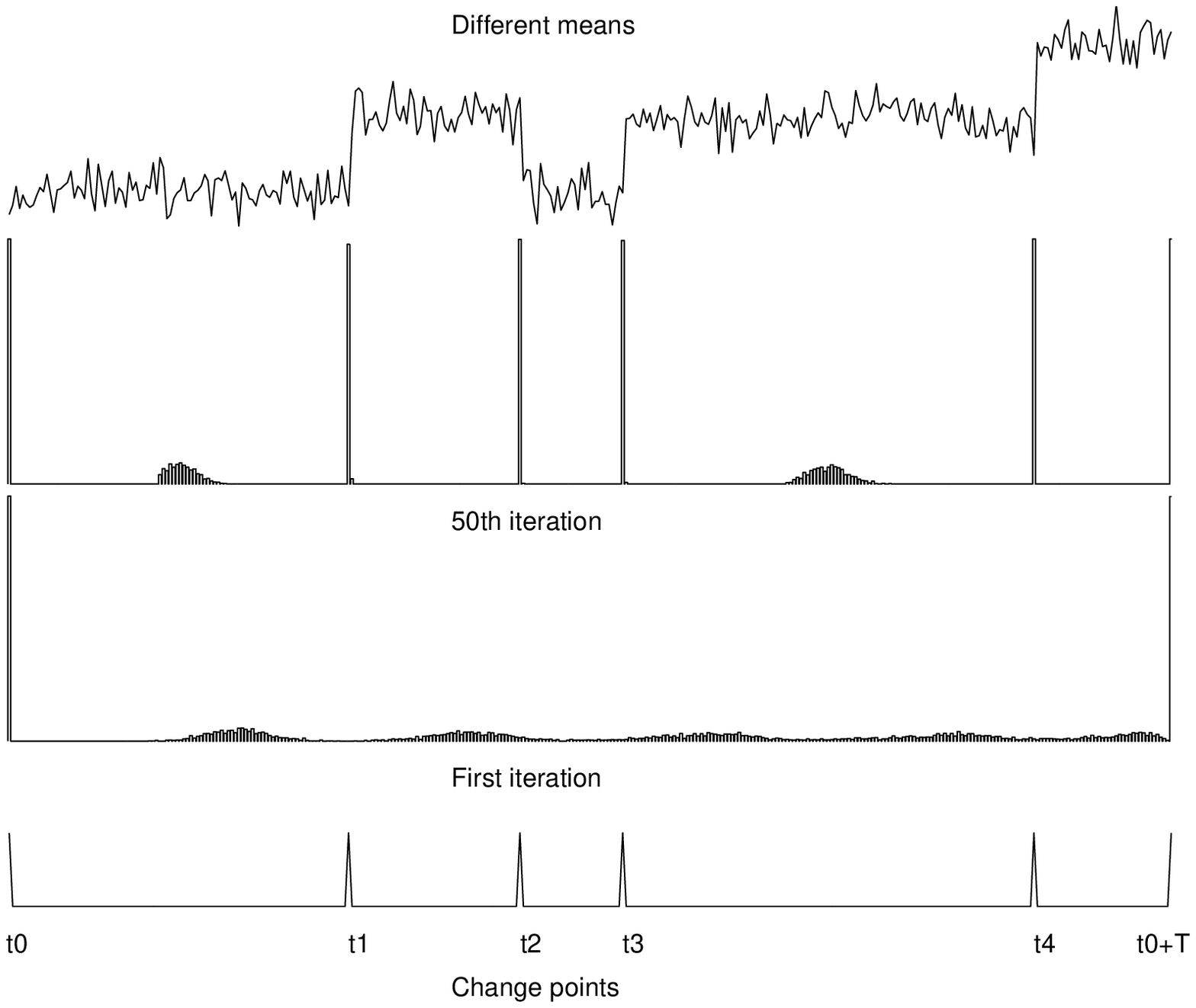} 
\ecc
\caption[Different means.]{Change in the means. up to down : simulated data, histogram in the 50th iteration, histogram in the first iteration, real position of the changepoints.}
\label{fig_mean}
\efig

\begin{table}[h]
\begin{center}
\begin{tabular}{|c|c|c|} \hline
m & $\hat{m}|\xb,\tb$ & $\hat{m}|\xb$ \\
\hline \hline
1.5 & 1.4966 & 1.4969  \\
1.7 & 1.7084 & 1.7013  \\
1.5 & 1.4912 & 1.5015  \\
1.7 & 1.6940 & 1.6929  \\
1.9 & 1.9012 & 1.8915  \\
\hline
\end{tabular}
\caption{Estimated value of the means}
\end{center}
\end{table}

\newpage
\subsection{Change in the variances}

We can  see in figure \ref{fig_var} that we have again good results on the position of the changepoints. However, for little difference of variances, the algorithm give an uncertainty on the exact position of the changepoint. This can be justified by the fact that the simulated data give itself this uncertainty. \\
In table 2 we can see again good estimations on the variances on each segments. 
\bfig[hbt]
\bcc
\includegraphics[width=150mm,height=75mm]{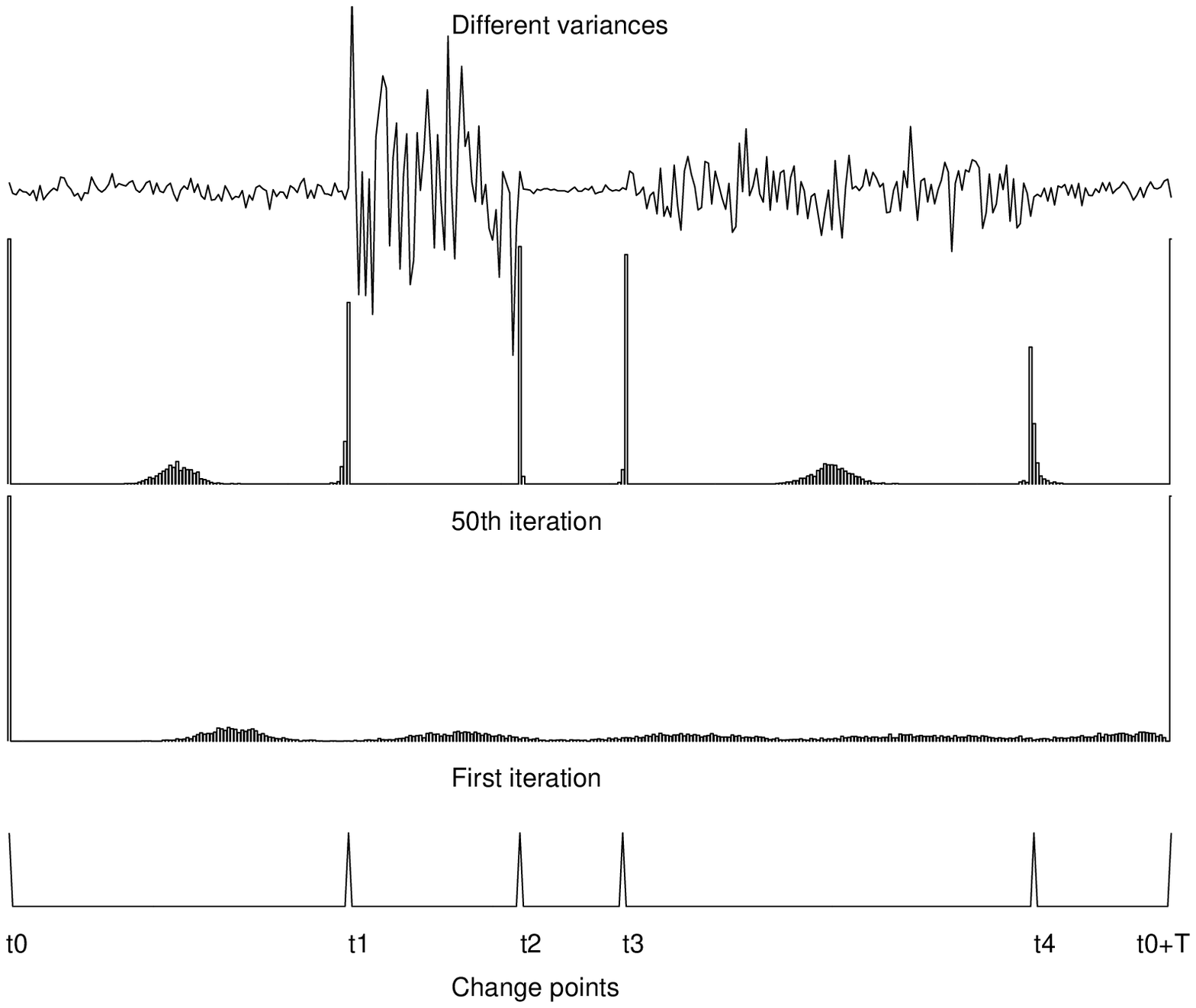} 
\ecc
\caption[Different variances.]{Change in the variances. up to down : simulated data, histogram in the 50th iteration, histogram in the first iteration, real position of the changepoints.}
\label{fig_var}
\efig

\begin{table}[h]
\begin{center}
\begin{tabular}{|c|c|c|} \hline
$\sigma^2$ & $\hat{\sigma}^2|\xb,\tb$ & $\hat{\sigma}^2 |\xb$ \\
\hline \hline
0.01 & 0.0083 & 0.0081 \\
1 & 0.9918 & 0.9598 \\
0.001 & 0.0007 & 0.0026 \\
0.1 & 0.0945 & 0.0940 \\
0.01 & 0.0079 & 0.0107 \\
\hline
\end{tabular}
\caption{Estimated value of the variances}
\end{center}
\end{table}

\newpage
\subsection{Change in the correlation coefficient}
The results showed in figure \ref{fig_corr_coef} are worse than in the two first cases. The position of the changepoints are less precise, and we can see that another changepoint appears. This affects the estimation of the correlation coefficient in the third segment because the algorithm alternates between two positions of changepoint. This problem can be justified by the fact that a value of the correlation coefficient near 1 implies locally a change of the mean, which can be considered by the algorithm as a changepoint. Also this problem appears when the size of the segments are far from the \aprio size $\lambda$. 
\bfig[hbt]
\bcc
\includegraphics[width=150mm,height=75mm]{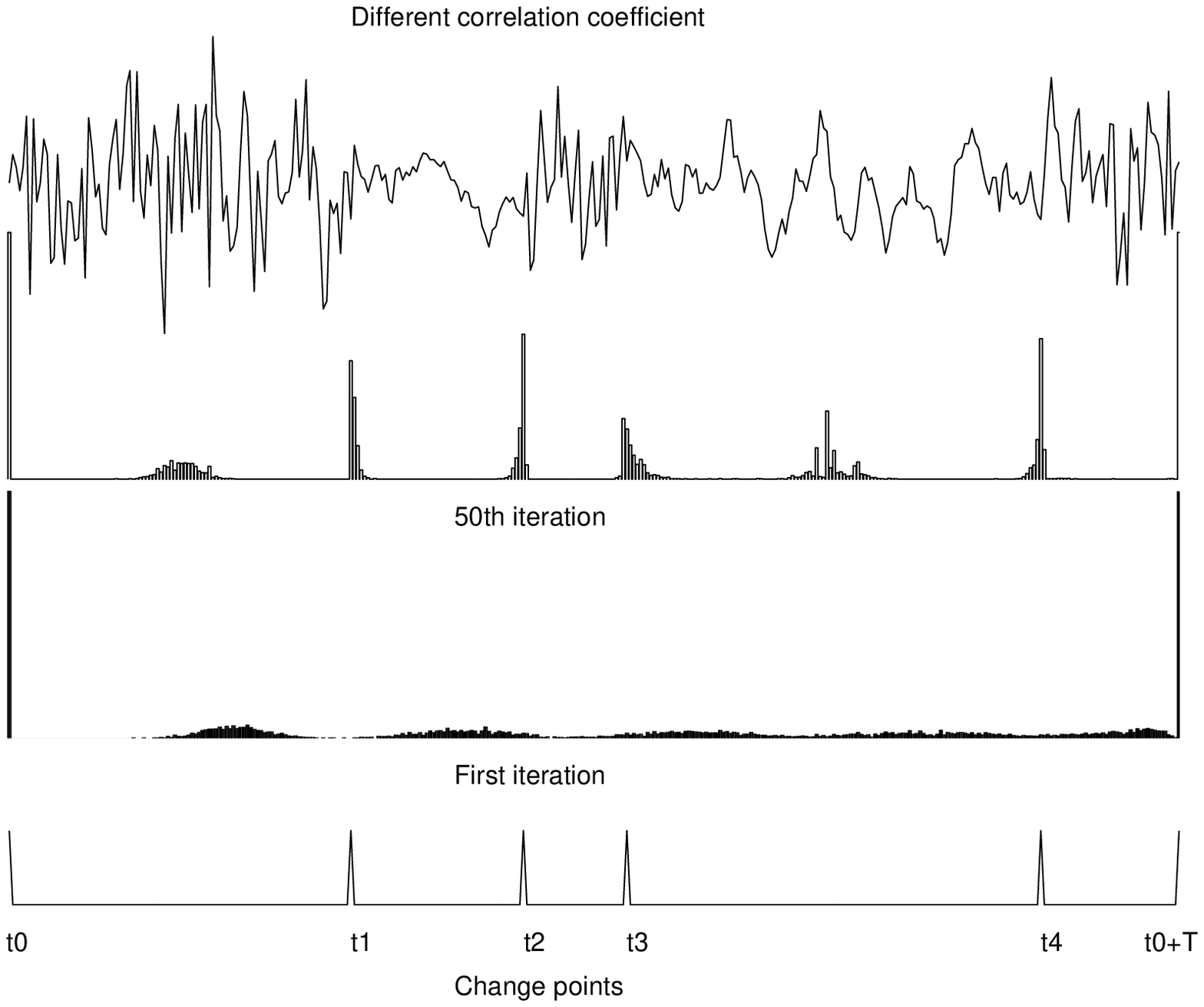} 
\ecc
\caption[Different correlation coefficients.]{Change in the correlation coefficient. up to down : simulated data, histogram in the 50th iteration, histogram in the first iteration, real position of the changepoints.}
\label{fig_corr_coef}
\efig

\begin{table}[h]
\begin{center}
\begin{tabular}{|c|c|} \hline
$a$  & $\hat{a}|\xb$ \\
\hline \hline
0 & 0.0988 \\
0.9 & 0.7875 \\
0.1 & 0.3737 \\
0.8 & 0.8071 \\
0.2 & 0.1710 \\
\hline
\end{tabular}
\caption{Estimated vaue of the correlation coefficients}
\end{center}
\end{table}

\newpage
\subsection{Influence of the prior law}
\noindent
In this section we study the influence of the \aprio on $\lambda$, \ie the size of the segments. In the following we fix the number of changepoints as before and we change the \aprio size of the segments by $\lambda_0=\frac{\lambda}{2}$ and $\lambda_1=2\lambda$. We apply then our algorithm on the change of the correlation coefficient.
\bfig[hbt]
\bcc
\includegraphics[width=150mm,height=75mm]{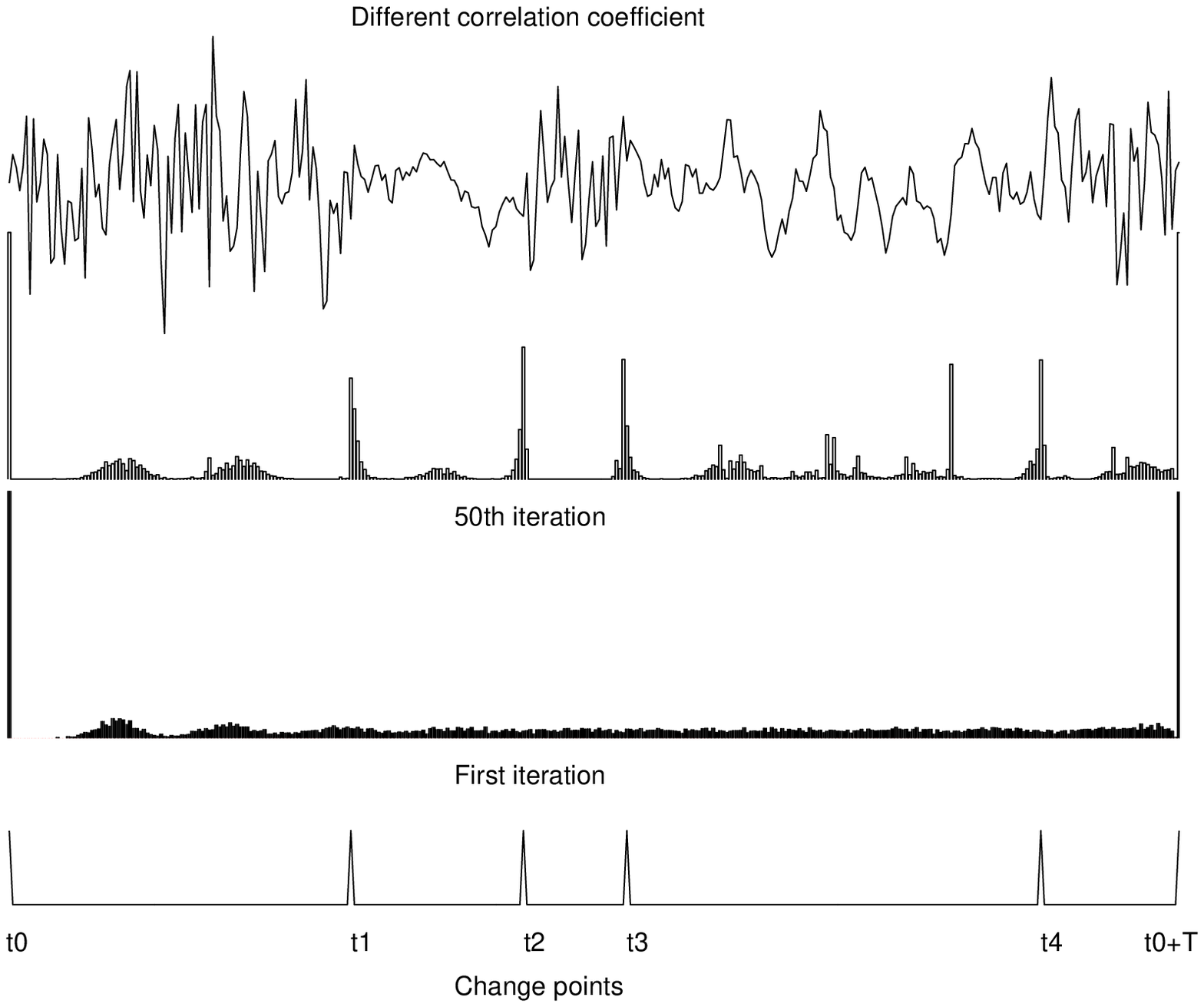} 
\ecc
\caption[Change in the correlation coefficients]{Different correlation coefficient with $\lambda_0=\frac{1}{2}\frac{T}{N+1}$. up to down : simulated data, histogram in the 50th iteration, histogram in the first iteration, real position of the changepoints.}
\label{fig_corr_coef_1}
\efig

\bfig[hbt]
\bcc
\includegraphics[width=150mm,height=75mm]{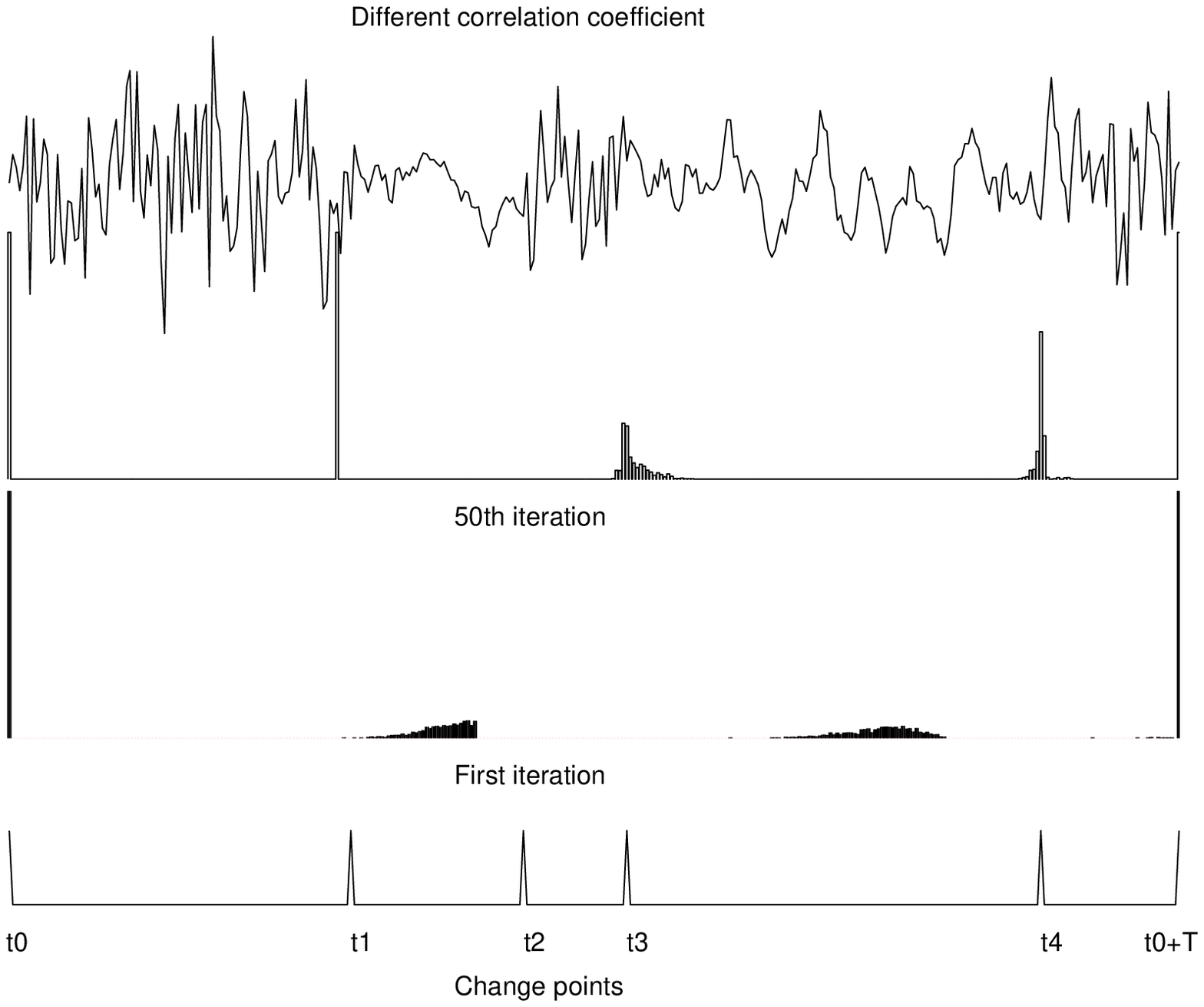} 
\ecc
\caption[Change in the correlation coefficients.]{Different correlation coefficient with $\lambda_1=2\frac{T}{N+1}$. up to down : simulated data, histogram in the 50th iteration, histogram in the first iteration, real position of the changepoints.}
\label{fig_corr_coef_2}
\efig

\noindent
In figure \ref{fig_corr_coef_1}, we can see that the algorithm has detected other changepoints, forming segments whose size is near $\lambda_0$. This result shows the importance of the \aprio when the data are not enough significant. We can also see this conclusion in figure \ref{fig_corr_coef_2} where only three changepoints are detected, forming segments whose size is again near $\lambda_1$. We can also remark that fixing \aprio a size $\lambda$ comes down to fix the number of changepoints. Our algorithm give then good results for instance if we have a good \aprio on the number of changepoints.

\section{Conclusions}

\small 
\bibliographystyle{ieeetr}
\bibliography{revuedef,biben,baseAJ,baseKZ,cp}

\begin{thebibliography}{10}

\bibitem{Basseville88}
M.~Basseville, ``Detecting changes in signals and systems -- a survey,'' {\em
  Automatica}, vol.~24, no.~3, pp.~309--326, 1988.

\bibitem{Wax91}
M.~Wax, ``Detection and localization of multiple sources via the stochastic
  signals model,'' {\em \uppercase{ieee} {T}ransactions on {S}ignal
  {P}rocessing}, vol.~39, pp.~2450--2456, {N}ovember 1991.

\bibitem{Kormylo82}
J.~J. Kormylo and J.~M. Mendel, ``Maximum-likelihood detection and estimation
  of {B}ernoulli-{G}aussian processes,'' {\em \uppercase{ieee} {T}ransactions
  on {I}nformation {T}heory}, vol.~28, pp.~482--488, 1982.

\bibitem{Chi85}
C.~Y. Chi, J.~Goustias, and J.~M. Mendel, ``A fast maximum-likelihood
  estimation and detection algorithm for {B}ernoulli-{G}aussian processes,'' in
  {\em {P}roceedings of the {I}nternational {C}onference on {A}coustic,
  {S}peech and {S}ignal {P}rocessing}, (Tampa, \sca{fl}), pp.~1297--1300,
  {A}pril 1985.

\bibitem{Goutsias88}
J.~K. Goutsias and J.~M. Mendel, ``Optimal simultaneous detection and
  estimation of filtered discrete semi-{M}arkov chains,'' {\em \uppercase{ieee}
  {T}ransactions on {I}nformation {T}heory}, vol.~34, pp.~551--568, 1988.

\bibitem{Oliver96}
J.~J. Oliver, R.~A. Baxter, and C.~S. Wallace, ``Unsupervised {L}earning using
  {MML},'' in {\em Machine Learning: Proceedings of the Thirteenth
  International Conference ({ICML} 96)}, pp.~364--372, Morgan Kaufmann
  Publishers, 1996.

\bibitem{Hughes99}
J.~P. Hughes, P.~Guttorp, and S.~P. Charles, ``A non-homogeneous hidden
  {Markov} model for precipitation occurrence,'' {\em Applied Statistics},
  vol.~48, no.~1, pp.~15--30, 1999.

\bibitem{Fitzgibbon00}
L.~J. Fitzgibbon, L.~, and D.~L. Dowe, ``Minimum message length grouping of
  ordered data,'' in {\em Algorithmic Learning Theory, 11th International
  Conference, {ALT} 2000, Sydney, Australia, December 2000, Proceedings},
  vol.~1968, pp.~56--70, Springer, Berlin, 2000.

\bibitem{Fitzgibbon02}
L.~Fitzgibbon, D.~L. Dowe, and L.~Allison, ``Change-point estimation using new
  minimum message length approximations,'' in {\em Proceedings of the Seventh
  Pacific Rim International Conference on Artificial Intelligence
  (PRICAI-2002)} (M.~Ishizuka and A.~Sattar, eds.), vol.~2417 of {\em LNAI},
  (Berlin), pp.~244--254, Japanese Society for Artificial Intelligence (JSAI),
  Springer-Verlag, {A}ugust 2002.

\bibitem{Fearnhead}
P.~Fearnhead, ``Exact and efficient bayesian inference for multiple changepoint
  problems,'' tech. rep., Department of math. and stat., Lancaster university.

\end{thebibliography}

\edoc